\documentclass[10pt,preprint,amsmath,amssymb]{revtex4}
\usepackage{fontenc}
\usepackage[latin1]{inputenc}
\usepackage{graphicx}

\newtheorem{defi}{Definition}[section]
\newtheorem{satz}{Theorem}[section]
\newtheorem{cor}{Corollar}[section]

\begin{document}

\title{On a generalization of Jacobi's elliptic functions and the Double Sine-Gordon kink chain}
\author{Michael Pawellek}
\affiliation{Institut für Theoretische Physik III, \\ Universität Erlangen-Nürnberg, \\ Staudtstr.7, D-91058 Erlangen, Germany}
\email{michi@theorie3.physik.uni-erlangen.de}

\begin{abstract}
A generalization of Jacobi's elliptic functions is introduced as inversions of hyperelliptic integrals.
We discuss the special properties of these functions, present addition theorems and give a list of indefinite integrals. 
As a physical application we show that periodic kink solutions (kink chains) of the double sine-Gordon model can be described in 
a canonical form in terms of generalized Jacobi functions.
\end{abstract}

\maketitle

\section{Introduction}
The inversion of the Abelian integral
\begin{equation}\label{eq:Abel}
 u=\int_{x_0}^x\frac{\mathrm{d}tR(t)}{\sqrt{P(t)}},
\end{equation}
where \(R(t)\) is a rational function and \(P(t)\) is a polynom of degree \(p\), is a problem, 
which has attracted many mathematicians e.g. Euler, Jacobi and, of course, Abel.
So for \(p=2\) and \(P(x)=(1-x^2)\) the inversion of (\ref{eq:Abel}) gives the periodic trigonometric functions \(x=\sin(u)\) and \(x=\cos(u)\). 
For \(p=4\) and \(P(x)=(1-x^2)(1-k^2x^2)\) (\ref{eq:Abel}) becomes an elliptic integral and its inversion leads to the doubly-periodic Jacobi 
elliptic functions \(x=\mathrm{sn}(u), x=\mathrm{cn}(u)\), etc.

For \(p>4\) the integral (\ref{eq:Abel}) is hyperelliptic and as was shown by Jacobi its inversion leads to infinite-valued functions with 
more than two independent periods \cite{Jaco}. In order to overcome infinite-valued functions  
Jacobi invented his celebrated inversion theorem, where the inversion of a system of Abelian integrals leads to one-valued functions of multiple
periodicity but depending on two or more independent variables \cite{Bake}. Therefore the inversion of a single Abelian integral, especially 
the hyperelliptic ones was only rarely considered by mathematicians \cite{Ludw, Groe} although simply mechanical problems can lead to these
integrals. The determination of the trajectory \(x(t-t_0)\) of a point particle in a potential \(V(x)\) given by a polynomial of degree greater
than four needs the inversion of the integral \cite{Fedo}:
\begin{equation}
 t-t_0=\int_{x_0}^x\frac{\mathrm{d}x'}{\sqrt{2(E-V(x'))}}.
\end{equation} 

There are special cases for \(p>4\) where the inversion of one single Abelian integral leads to 
multi-valued functions, e.g. when one can reduce the hyperelliptic integral to an elliptic one.

In this work we will consider a set of functions \(\{s(u),c(u),d_1(u),d_2(u)\}\), which are inversions of certain hyperelliptic integrals, where
\(P(x)\) are polynomials of degree 6. As we will show, they can be understood as generalizations of the Jacobi elliptic functions of  
the case \(p=4\). For example, the relation
\begin{equation}
 \mathrm{sn}'(u)=\mathrm{cn}(u)\mathrm{dn}(u)
\end{equation}
will be extended to
\begin{equation}
 s'(u)=c(u)d_1(u)d_2(u).
\end{equation}
This generalization of Jacobi's elliptic functions were recently identified as special solutions of a generalization of Lam\'e's 
differential equation \cite{Pawe} and in the following sections we will continue to discuss the mathematical properties of these functions. 

As a physical application of 
these functions we will show, that several periodic kink solutions of the double sine-Gordon model \cite{Iwab, Huda, Wang}, which are expressible 
as nested combinations of text book functions are just reincarnations of one unique generalized Jacobi function.

\section{Definitions}
In this section we introduce the generalized Jacobi functions and clarify some of their properties, which were already used in \cite{Pawe}.
\begin{defi}
Consider without loss of generality \(1>k_1>k_2>0\) as moduli parameter. 

a) The generalized Jacobi elliptic function \(x=s(u,k_1,k_2)\) and their companion functions \(c(u,k_1,k_2), d_1(u,k_1,k_2)\) and 
\(d_2(u,k_1,k_2)\)
are defined by the inversion of the hyperelliptic integrals
\begin{eqnarray}\label{eq:Hyperell}
 u(x,k_1,k_2)&=&\int_0^{x=s(u)}\frac{\mathrm{d}t}{\sqrt{(1-t^2)(1-k_1^2t^2)(1-k_2^2t^2)}},\\
 u(x,k_1,k_2)&=&\int_{x=c(u)}^1\frac{\mathrm{d}t}{\sqrt{(1-t^2)({k'_1}^2+k_1^2t^2)({k'_2}^2+k_2^2t^2)}},\\
 u(x,k_1,k_2)&=&k_1\int_{x=d_1(u)}^1\frac{\mathrm{d}t}{\sqrt{(1-t^2)(t^2-{k'_1}^2)(k_1^2-k_2^2+k_2^2t^2)}},\\
 u(x,k_1,k_2)&=&k_2\int_{x=d_2(u)}^1\frac{\mathrm{d}t}{\sqrt{(1-t^2)(t^2-{k'_2}^2)(k_2^2-k_1^2+k_1^2t^2)}},
\end{eqnarray}
respectively.

b) The generalized amplitude function \(a(u,k_1,k_2)\) is the given by the inversion of  
\begin{equation}
 u(\varphi_,k_1,k_2)=\int_0^{\varphi=a(u)}\frac{\mathrm{d}\psi}{\sqrt{(1-k_1^2\sin^2\psi)(1-k_2^2\sin^2\psi)}}
\end{equation}
with \(s(u,k_1,k_2)=\sin(a(u,k_1,k_2))\).

\end{defi}

Without solving the integrals (\ref{eq:Hyperell}) explicitly, one can derive certain properties of these functions. 

\begin{cor}
Given the generalized Jacobi elliptic functions \(s(u), c(u), d_1(u)\) and \(d_2(u)\) as defined by (\ref{eq:Hyperell}).
Then 
\begin{equation}\label{eq:Companion}
 c^2(u)=1-s^2(u),\qquad d_1^2(u)=1-k_1^2s^2(u),\qquad d_2^2(u)=1-k_2^2s^2(u),
\end{equation}
\begin{equation}\label{eq:Rel1}
 d_i^2(u)-k_i^2c^2(u)=1-k_i^2,\;i=1,2;\qquad k_1^2d_2^2(u)-k_2^2d_1^2(u)=k_1^2-k_2^2.
\end{equation}

The first derivatives of these functions are given by 
\begin{eqnarray}\label{eq:firstderiv}
 s'(u)=c(u)d_1(u)d_2(u),\qquad c'(u)=-s(u)d_1(u)d_2(u),\nonumber\\
 d_1'(u)=-k_1^2s(u)c(u)d_2(u),\qquad d_2'(u)=-k_2^2s(u)c(u)d_1(u),
\end{eqnarray}

\end{cor}

\paragraph*{\bf Proof}
By the substitutions 
\begin{equation}
 x=\sqrt{1-y^2},\qquad x=\sqrt{1-k_1^2y^2},\qquad x=\sqrt{1-{k_2}^2y^2}
\end{equation}
in (\ref{eq:Hyperell}) one obtains the three other integrals where the relations (\ref{eq:Companion}) and (\ref{eq:Rel1}) 
can be read off. Relation (\ref{eq:firstderiv}) follows from the differential versions of (\ref{eq:Hyperell}). \(\Box\)

The functions  \(s(u,k_1,k_2), c(u,k_1,k_2), d_1(u,k_1,k_2)\) and \(d_2(u,k_1,k_2)\) are generalizations 
of the classic Jacobi elliptic functions \(\mathrm{sn}(u,k),\mathrm{cn}(u,k)\) and \(\mathrm{dn}(u,k)\) 
and they reduce to them for \(k_2\to 0\) and \(k=k_1\). For fixed \(k_i\) we will us the abbreviated notations \(s(u)\equiv s(u,k_1,k_2)\), etc.

So far we have only stated some formal relations between the inverted hyperelliptic integrals 
(\ref{eq:Hyperell}), provided these function exist, which we have to show next. 
For this we note that the differential of the hyperelliptic integral, which defines \(s(u)\) is an Abelian differential 
of the first kind

\begin{figure}\label{fig:genjacobis2}
 \includegraphics[scale=0.6]{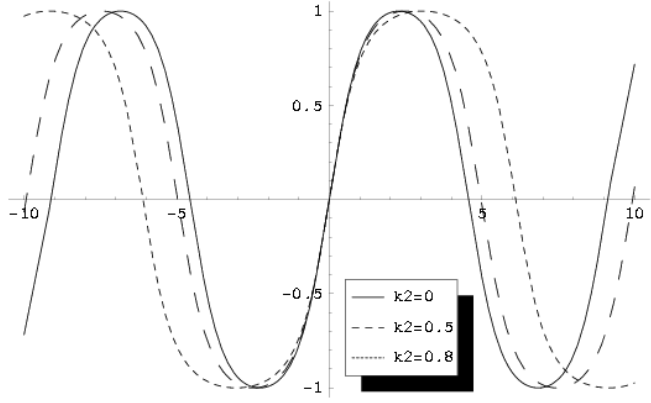}
 \caption{\(s(x,k_1,k_2)\) for \(k_1=0.9\)}
\end{figure}
\begin{figure}\label{fig:genjacobis3}
 \includegraphics[scale=0.6]{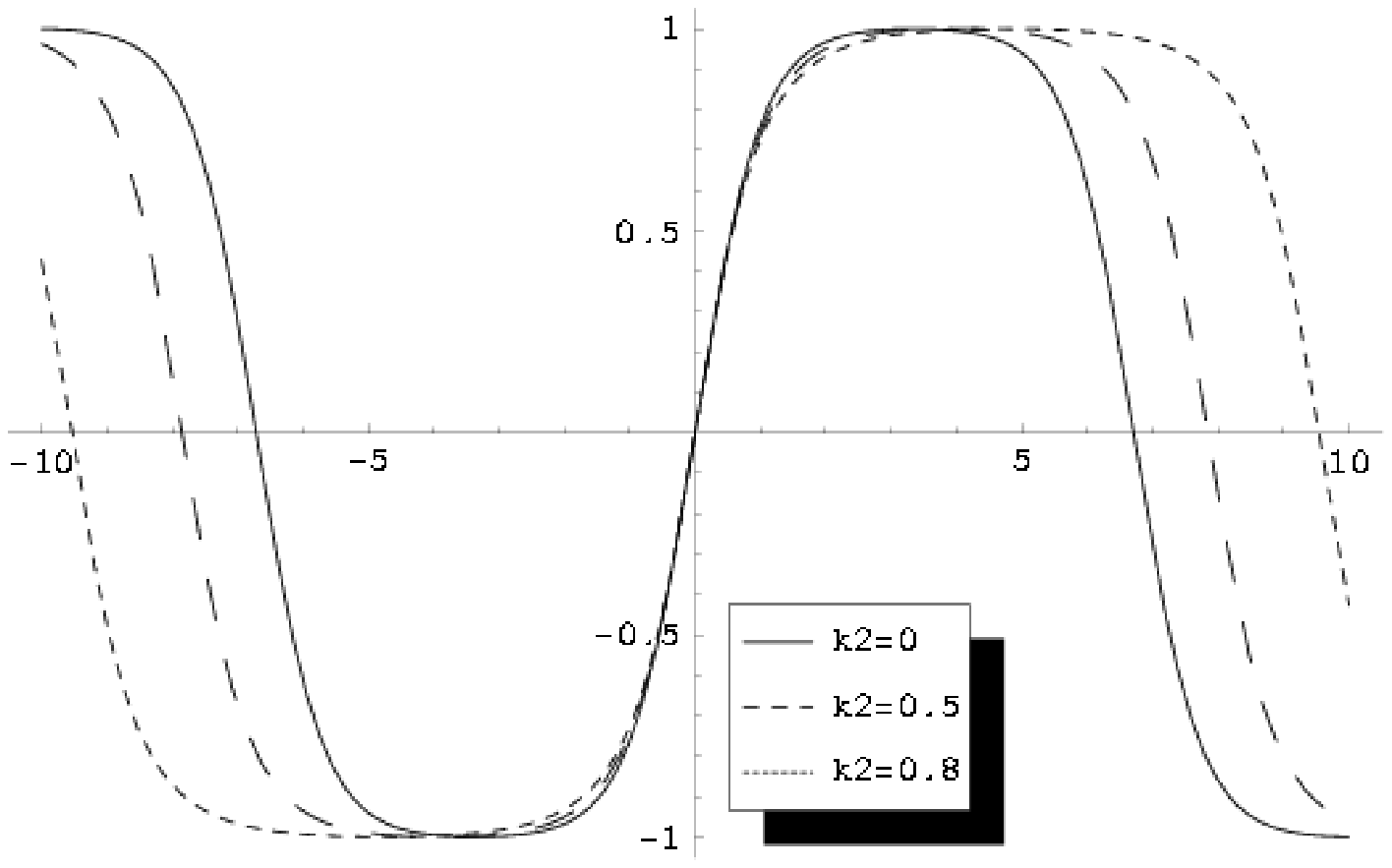}
 \caption{\(s(x,k_1,k_2)\) for \(k_1=0.99\)}
\end{figure}

\begin{figure}\label{fig:genjacobic3}
 \includegraphics[scale=0.6]{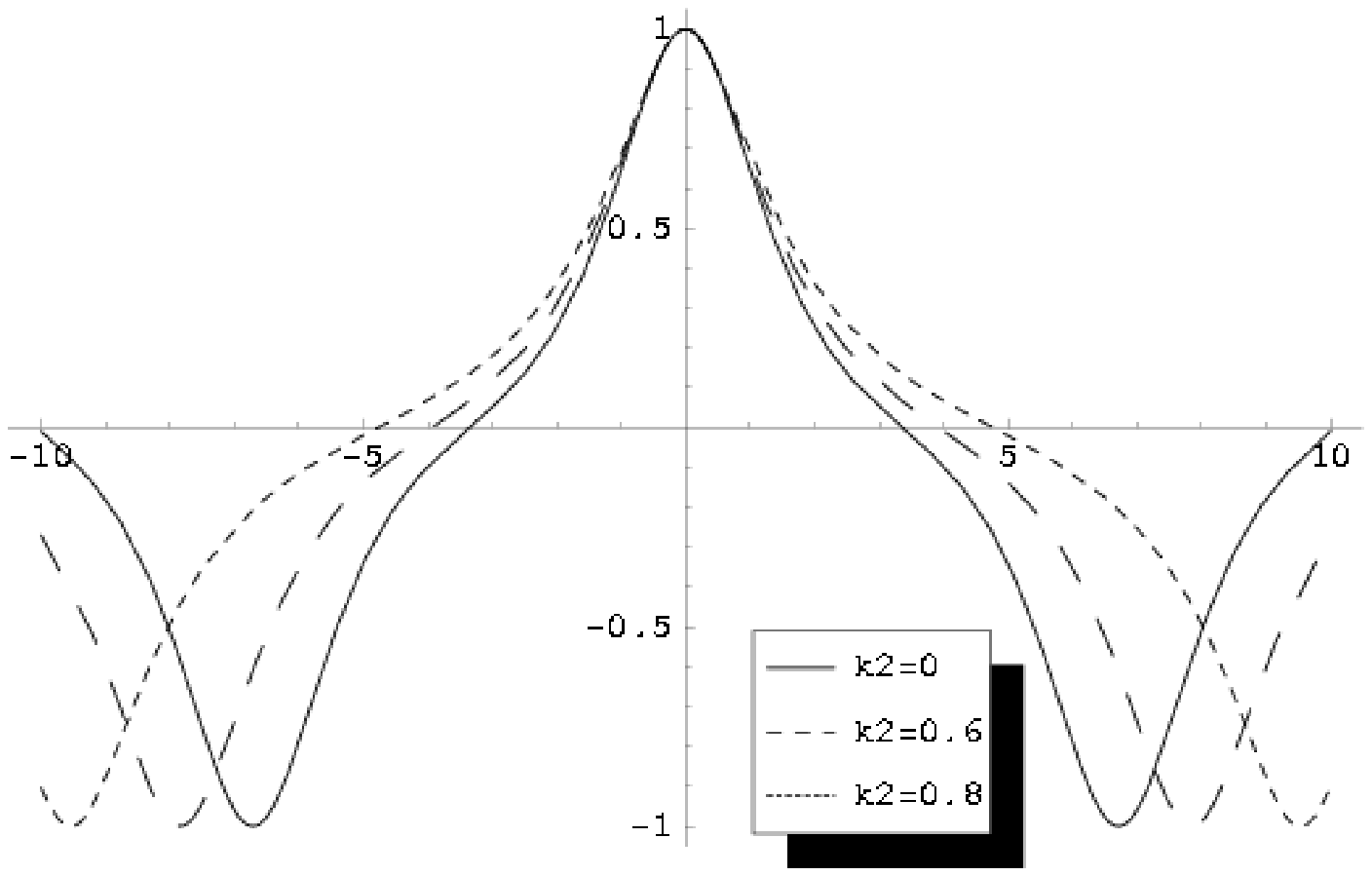}
 \caption{\(c(x,k_1,k_2)\) for \(k_1=0.99\)}
\end{figure}

\begin{figure}\label{fig:genjacobid23}
 \includegraphics[scale=0.6]{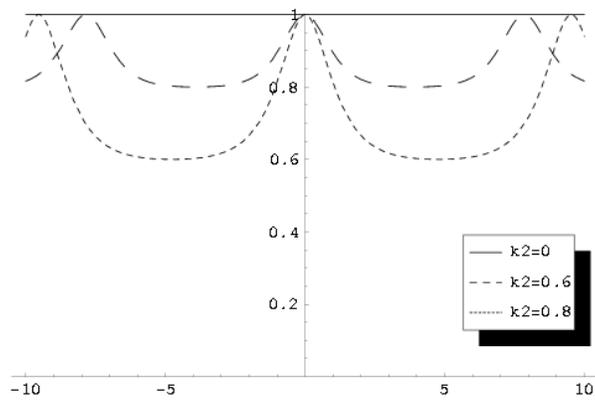}
 \caption{\(d_2(x,k_1,k_2)\) for \(k_1=0.99\)}
\end{figure} 

\begin{equation}\label{eq:Abeldiff}
 \mathrm{d}\eta=\frac{\mathrm{d}x}{y}
\end{equation}
with 
\begin{equation}
 y^2=(1-x^2)(1-k_1^2x^2)(1-k_2^2x^2)
\end{equation}
It is holomorphic on the hyperelliptic curve \(\mathcal{C}\) defined by
\begin{equation}
 \mathcal{C}=\{(y,x)\in\mathrm{C}^2 | y^2=(1-x^2)(1-k_1^2x^2)(1-k_2^2x^2)\},
\end{equation}
which can be modelled as a Riemann surface of genus 2.
The important observation is that the hyperelliptic curve \(\mathcal{C}\) is also a double cover 
\(\mathcal{C}\stackrel{\pi}{\longrightarrow}\mathcal{E}\) of the elliptic curve \(\mathcal{E}\) defined by
\begin{equation}
 \mathcal{E}=\{(w,z)\in\mathrm{C}^2 | w^2=z(1-z)(1-k_1^2z)(1-{k_2}^2z)\},
\end{equation}
with covering map \(\pi(y,x)\) given by
\begin{equation}
 (w,z)=\pi(y,x)=(xy,x^2).
\end{equation}
The differential (\ref{eq:Abeldiff}) is therefore the pullback of the elliptic differential of the first kind 
\begin{equation}
 \mathrm{d}\eta=\frac{\mathrm{d}z}{w},
\end{equation}
and the inversion of its integral gives a double-valued function, which can now be expressed in terms of elliptic functions:

\begin{satz} The generalized Jacobi elliptic functions exist and are given by
 \begin{eqnarray}\label{eq:Solution}
 s(u,k_1,k_2)=\frac{\mathrm{sn}(k_2'u,\kappa)}{\sqrt{{k'_2}^2+k_2^2\mathrm{sn}^2(k_2'u,\kappa)}},\qquad
 c(u,k_1,k_2)=\frac{k_2'\mathrm{cn}(k_2'u,\kappa)}{\sqrt{1-k_2^2\mathrm{cn}^2(k_2'u,\kappa)}},\nonumber\\ \nonumber\\
 d_1(u,k_1,k_2)=\frac{\sqrt{k_1^2-k_2^2}\mathrm{dn}(k_2'u,\kappa)}{\sqrt{k_1^2-k_2^2\mathrm{dn}^2(k_2'u,\kappa)}},\qquad
 d_2(u,k_1,k_2)=\frac{\sqrt{k_1^2-k_2^2}}{\sqrt{k_1^2-k_2^2\mathrm{dn}^2(k_2'u,\kappa)}},
\end{eqnarray}
and the generalized amplitude function is
\begin{equation}
 a(u,k_1,k_2)=\arctan[{k'_2}^{-1}\mathrm{sc}(k'_2u,\kappa)]=\arctan[{k'_2}^{-1}\tan(\mathrm{am}(k'_2u,\kappa))]
\end{equation}
with \(\kappa^2=(k_1^2-k_2^2)/(1-k_2^2)\), \(k_2'=\sqrt{1-k_2^2}\) and \(0\leq k_2\leq k_1\leq 1\).
They have branch-cuts along \((u_1,u_2)\) and \((u_3,u_4)\) with 
\begin{eqnarray}
 u_1=i\frac{\mathrm{cn}^{-1}(k_2,\kappa')}{k_2'},\qquad u_2=-u_1+2i\frac{\mathbf{K}(\kappa')}{k_2'},\qquad
 u_3=u_1+2\frac{\mathbf{K}(\kappa)}{k_2'}, \qquad u_4=u_2+2\frac{\mathbf{K}(\kappa)}{k_2'},
\end{eqnarray}
where \(\mathbf{K}(k)\) is the complete elliptic integral of the first kind and \(\kappa'=\sqrt{1-\kappa^2}\).
\end{satz}

\paragraph*{\bf Proof} From the discussion above follows that by substituting \(t=\sqrt{\tau}\), the hyperelliptic integral (\ref{eq:Hyperell}) 
can be reduced to the following elliptic integral:
\begin{equation}
 u(x,k_1,k_2)=\frac{1}{2}\int_0^{x^2}\frac{\mathrm{d}\tau}{\sqrt{\tau(1-\tau)(1-k_1^2\tau)(1-k_2^2\tau)}}, 
\end{equation} 
where the inverse function is given \cite{Byrd} by the first expression of (\ref{eq:Solution}). The sign of the root in the denominator is chosen
in such a way that for \(k_2\to 0\) one has \(s(u,k_1,k_2)\to \mathrm{sn}(u,k_1)\). The other three expressions are obtained by
applying (\ref{eq:Companion}). The branch points are a result of the zeros of the denominators in (\ref{eq:Solution}). \(\Box\)

Figures 1 to 4 show example plots of these functions for selected values of the moduli \(k_1\) and \(k_2\). 

As the Jacobi representation (\ref{eq:Solution}) shows, the introduction of generalized Jacobi functions is mathematically redundant.
Nevertheless it would be not obvious in the Jacobi representation that among these four functions such elementary relations as 
(\ref{eq:firstderiv}) are fulfilled.
It is therefore advantageous to use (\ref{eq:Companion}) to (\ref{eq:firstderiv}) when working with these functions and not representation 
(\ref{eq:Solution}). With this set-up algebraic manipulations become very simply and straight-forward.

Further, the generalized Jacobi functions serve as prototype examples of meromorphic functions on a genus two Riemann surface. This can be
seen as follows.
Consider the two points \(u_1=u\) and \(u_2=u+2\frac{\mathbf{K}(\kappa)}{k'_2}\). There exist two different paths for analytic continuation
to obtain the value of \(s(u_2)\) from \(s(u_1)\). Path \(a_1\) avoids the branch cut and path \(a_2\) goes through one cut, see Figure 5. 
After passing the cut \((u_1,u_2)\) one has to use the other branch of the square root. Let \((u,+)\) and \((u,-)\) denote points lying
in the two different branches of the square root. Then one gets
\begin{equation}
 s((u,-))=\frac{\mathrm{sn}(k_2'u)}{-\sqrt{{k'_2}^2+k_2^2\mathrm{sn}^2(k'_2u)}}=
 \frac{\mathrm{sn}(k_2'u+2\mathbf{K})}{\sqrt{{k'_2}^2+k_2^2\mathrm{sn}^2(k'_2u+2\mathbf{K})}}=s((u+2\mathbf{K}(\kappa)/k'_2,+)), 
\end{equation}
where we have used the anti-periodicity of the \(\mathrm{sn}\)-function.
Thus by identifying the points \((u,-)\sim (u+2\mathbf{K}(\kappa)/k'_2,+)\) of the two branches, the path \(a_2\) enters the cut \((u_1,u_2)\) and 
appears at the other cut \((u_3,u_4)\). The branch cuts are short cuts and depending on the path of analytic continuation one gets:
\begin{equation}
 s(u+4\mathbf{K}(\kappa)/k'_2)_{a_1}=s(u),\qquad s(u+2\mathbf{K}(\kappa)/k'_2)_{a_2}=s(u).
\end{equation}
Thus the generalized Jacobi functions are realizations of functions with two real periods \(2\mathbf{K}(\kappa)/k'_2\) and
\(4\mathbf{K}(\kappa)/k'_2\), depending on the path of analytic continuation. The identification of the nontrivial cycles as in Figure 5
makes it clear that the generalized Jacobi functions are one-valued functions on the corresponding genus two Riemann surface. The cycles
\(b_1\) and \(b_2\) in Figure 5 correspond to the imaginary period \(2i\mathbf{K}(\kappa')/k'_2\). One can think
of this surface as a torus with an additional handle attached connecting the branch cuts.

\begin{figure}
 \includegraphics[scale=0.7]{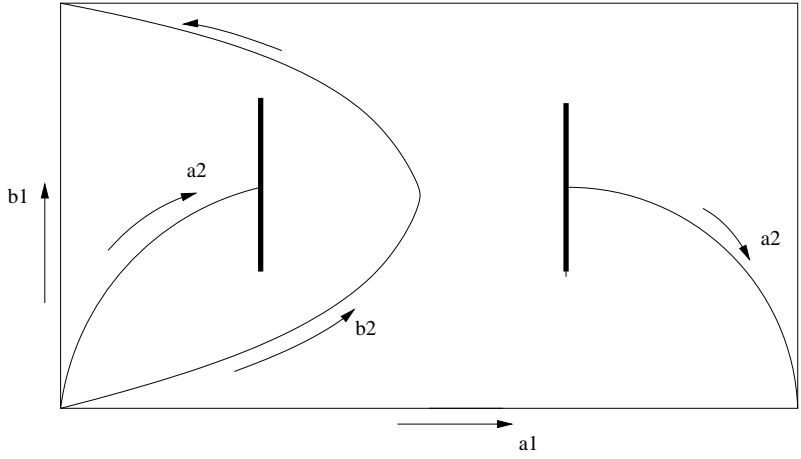}
 \caption{A fundamental cell in the complex plane with 'short cuts' and the four periods}
\end{figure}

\section{Properties}
In this section we present addition theorems, special values and indefinite integrals of the generalized Jacobi elliptic functions.

\subsection{Relation to classic Jacobi elliptic functions}

From (\ref{eq:Solution}) we can state the following
\begin{cor}
The 12 classic Jacobi elliptic functions are given by the nontrivial quotients of the generalized Jacobi functions, e.g. one has
\begin{equation}\label{eq:ratio}
 \frac{s(u,k_1,k_2)}{d_2(u,k_1,k_2)}=k_2'^{-1}\mathrm{sn}(k_2'u,\kappa),\;\;\frac{c(u,k_1,k_2)}{d_2(u,k_1,k_2)}=\mathrm{cn}(k_2'u,\kappa),\;\;
 \frac{d_1(u,k_1,k_2)}{d_2(u,k_1,k_2)}=\mathrm{dn}(k_2'u,\kappa),
\end{equation}
where the modulus of the resulting Jacobi elliptic functions is \(\kappa\).
For the remaining nine quotients see Table I.
\end{cor}

This looks very similar to the definition of the Jacobi functions by theta functions \cite{Whit}: 
\begin{equation}\label{eq:ThetaJacobi}
 \mathrm{sn}(u)=\frac{\vartheta_3}{\vartheta_2}\frac{\vartheta_1(u/\vartheta_3^2)}{\vartheta_4(u/\vartheta_3^2)},\;\;
 \mathrm{cn}(u)=\frac{\vartheta_4}{\vartheta_2}\frac{\vartheta_2(u/\vartheta_3^2)}{\vartheta_4(u/\vartheta_3^2)},\;\;
 \mathrm{dn}(u)=\frac{\vartheta_4}{\vartheta_3}\frac{\vartheta_3(u/\vartheta_3^2)}{\vartheta_4(u/\vartheta_3^2)},
\end{equation}
where \(\vartheta_i=\vartheta_i(0)\). More similarity with theta functions can be found, when one notice that 
 from (\ref{eq:firstderiv}) especially follows the identity 
\begin{equation}
 s'(0)=c(0)d_1(0)d_2(0),
\end{equation}
which is also very similar to the famous theta constant identity \cite{Whit}
\begin{equation}
  \vartheta'_1(0)=\vartheta_2(0)\vartheta_3(0)\vartheta_4(0).
\end{equation}
\begin{table}\label{tab:ratio}
\caption{The ratios of generalized Jacobi functions with moduli \(k_1\) and \(k_2\) give the twelve Jacobi elliptic functions with modulus \(\kappa\)}
\begin{tabular}{c||c|c|c|c}
\hline
   & \(s(u)\) & \(c(u)\) & \(d_1(u)\) & \(d_2(u)\)  \\
 \hline\hline
 \(s(u)\) & \(1\) & \(k_2'\mathrm{cs}(k_2'u)\) & \(k_2'\mathrm{ds}(k_2'u)\) & \(k_2'\mathrm{ns}(k_2'u)\) \\ & & & & \\ 
 \(c(u)\) & \(k_2'^{-1}\mathrm{sc}(k_2'u)\) & \(1\) & \(\mathrm{dc}(k_2'u)\) & \(\mathrm{nc}(k_2'u)\) \\ & & & & \\
 \(d_1(u)\) & \(k_2'^{-1}\mathrm{sd}(k_2'u)\) & \(\mathrm{cd}(k_2'u)\) & \(1\) & \(\mathrm{nd}(k_2'u)\) \\ & & & & \\
 \(d_2(u)\) & \(k_2'^{-1}\mathrm{sn}(k_2'u)\) & \(\mathrm{cn}(k_2'u)\) & \(\mathrm{dn}(k_2'u)\) & \(1\) \\
\end{tabular}
\end{table}

Nevertheless a similar relation as (\ref{eq:firstderiv}) does not hold for theta functions:
\begin{equation}
 \vartheta'_1(u)\neq\vartheta_2(u)\vartheta_3(u)\vartheta_4(u),
\end{equation}
 which is a crucial difference to the generalized Jacobi functions.

\subsection{Addition theorems}

\begin{satz}[Addition theorem]
The generalized Jacobi functions with moduli \(k_2,k_2\) fulfill the following addition theorems:
\begin{eqnarray}
 s(u\pm v)=\frac{s(u)d_2(u)c(v)d_1(v) \pm s(v)d_2(v)c(u)d_1(u)}{\sqrt{[d_2^2(u)d_2^2(v)-\kappa^2k_2'^4s^2(u)s^2(v)]^2+k_2^2
  [s(u)d_2(u)c(v)d_1(v)\pm s(v)d_2(v)c(u)d_1(u)]^2}}\nonumber\\ \nonumber\\
 c(u\pm v)=\frac{c(u)d_2(u)c(v)d_2(v)\mp k_2'^2s(u)d_1(u)s(v)d_1(v)}{\sqrt{[d_2^2(u)d_2^2(v)-\kappa^2k_2'^4s^2(u)s^2(v)]^2+k_2^2
  [s(u)d_2(u)c(v)d_1(v)\pm s(v)d_2(v)c(u)d_1(u)]^2}}\nonumber\\ \nonumber\\
 d_1(u\pm v)=\frac{d_1(u)d_2(u)d_1(v)d_2(v)\mp\kappa^2{k'_2}^2s(u)c(u)s(v)c(v)}{\sqrt{[d_2^2(u)d_2^2(v)-\kappa^2k_2'^4s^2(u)s^2(v)]^2+k_2^2
  [s(u)d_2(u)c(v)d_1(v)\pm s(v)d_2(v)c(u)d_1(u)]^2}}\nonumber\\ \nonumber\\
 d_2(u\pm v)=\frac{d_2^2(u)d_2^2(v)-\kappa^2k_2'^4s^2(u)s^2(v)}{\sqrt{[d_2^2(u)d_2^2(v)-\kappa^2k_2'^4s^2(u)s^2(v)]^2+k_2^2
  [s(u)d_2(u)c(v)d_1(v)\pm s(v)d_2(v)c(u)d_1(u)]^2}}\nonumber
\end{eqnarray}
\end{satz}
\paragraph*{\bf Proof}
Write the addition theorem for \(\mathrm{sn}(u)\) with the help of (\ref{eq:ratio}) as
\begin{equation}\label{eq:adsn}
 \mathrm{sn}(k_2'u\pm k_2'v,\kappa)=k_2'\frac{s(u)d_2(u)c(v)d_1(v)\pm s(v)d_2(v)c(u)d_1(u)}{d_2^2(u)d_2^2(v)-\kappa^2k_2^4s^2(u)s^2(v)}
\end{equation}
The addition theorem for \(d_2(u)\) follows then immediately by using (\ref{eq:adsn}) in 
\begin{equation}
 d_2^2(u\pm v)=\frac{k_2'^2}{k_2'^2+k_2^2\mathrm{sn}^2(k_2'u\pm k_2'v,\kappa)}.
\end{equation}
Now one can use the addition theorem for \(d_2(u)\) in order to get the corresponding theorem for \(s(u)\) from
\begin{equation}
 s(u\pm v)=k_2'^{-1}\mathrm{sn}(k_2'u\pm k_2'v,\kappa)d_2(u\pm v),
\end{equation}
and similar for \(c(u)\) and \(d_1(u)\). \(\Box\)

A special case of the addition theorems is the following

\begin{cor}[Half argument]
\begin{eqnarray}
 s^2(u/2)&=&\frac{d_2(u)-c(u)}{d_2(u)-k_2^2c(u)+{k'_2}^2d_1(u)},\\
 c^2(u/2)&=&{k'_2}^2\frac{c(u)+d_1(u)}{d_2(u)-k_2^2c(u)+{k'_2}^2d_1(u)},\\
 d_1^2(u/2)&=&(k_1^2-k_2^2)\frac{c(u)+d_1(u)}{k_1^2d_2(u)-k_2^2d_1(u)+(k_1^2-k_2^2)c(u)},\\
 d_2^2(u/2)&=&(k_1^2-k_2^2)\frac{c(u)+d_2(u)}{k_1^2d_2(u)-k_2^2d_1(u)+(k1^2-k_2^2)c(u)}.
\end{eqnarray}
\end{cor}

\subsection{Special Values}
\begin{defi} The generalization of the complete elliptic integral of the first kind is 
\begin{equation}
 \mathcal{K}:=\mathcal{K}(k_1,k_2)=\frac{1}{k'_2}\mathbf{K}(\kappa)=\int_0^1\frac{\mathrm{d}t}{\sqrt{(1-t^2)(1-k_1^2t^2)(1-k_2^2t^2)}}.
\end{equation}
Define also \(\mathcal{K'}=\mathcal{K}(\kappa')\) and \(\kappa'^2=1-\kappa^2=\frac{1-k_1^2}{1-k_2^2}\).
\end{defi}

From the definition of the generalized Jacobi functions follows
\begin{equation}
 s(\mathcal{K})=1,\;\;c(\mathcal{K})=0,\;\;d_1(\mathcal{K})=k'_1,\;\;d_2(\mathcal{K})=k'_2.
\end{equation}
In Table II we summarize analytic expressions for the generalized Jacobi functions evaluated at specific points. 
As an example we will demonstrate that \(s(\mathcal{K}/2)=(1+k_1'k_2')^{-\frac{1}{2}}\). For this we choose \(u=v=\mathcal{K}/2\) in
the addition theorem for \(c(u+v)\). One gets
\begin{equation}
 c^2(\mathcal{K}/2)d_2^2(\mathcal{K}/2)-k_2'^2s^2(\mathcal{K}/2)d_1^2(\mathcal{K}/2)=0.
\end{equation}
This can be written as 
\begin{equation}
 (k_2^2+k_1^2k_2'^2)s^4(\mathcal{K}/2)-2s^2(\mathcal{K}/2)+1=0,
\end{equation}
with solution
\begin{equation}
 s(\mathcal{K}/2)=\pm\sqrt{\frac{1\pm k_1'k_2'}{k_2^2+k_1'k_2'^2}}.
\end{equation}
Considering the limit \(k_2\to 0\) one has to obtain the result \(\mathrm{sn}(\mathbf{K}/2)=(1+k_1')^{-\frac{1}{2}}\), which
fixes the signs such as
\begin{equation}
 s(\mathcal{K}/2)=+\sqrt{\frac{1-k_1'k_2'}{k_2^2+k_1^2k_2'^2}}.
\end{equation}
By writing the denominator as
\begin{equation}
 k_2^2+k_1^2k_2'^2=1-k_2'^2+k_1^2k_2'^2=1-k_1'^2k_2'^2=(1+k_1'k_2')(1-k_1'k_2'),
\end{equation}
the promised result \(s(\mathcal{K}/2)=(1+k'_1k'_2)^{-\frac{1}{2}}\) is obtained.
\begin{table}\label{tab:spec}
\caption{Special values for generalized Jacobi functions}
\begin{tabular}{c||c|c|c|c}
\hline
  & \(s(u)\) & \(c(u)\) & \(d_1(u)\) & \(d_2(u)\) \\
 \hline\hline & & & & \\
\(u=\mathcal{K}/2\) & \((1+k'_1k'_2)^{-\frac{1}{2}}\) & \((k'_1k'_2)^{\frac{1}{2}}(1+k'_1k'_2)^{-\frac{1}{2}}\) &
 \({k'_1}^{\frac{1}{2}}(k'_1+k'_2)^{\frac{1}{2}}(1+k'_1k'_2)^{-\frac{1}{2}}\) & \({k'_2}^{\frac{1}{2}}(k'_1+k'_2)^{\frac{1}{2}}(1+k'_1k'_2)^{-\frac{1}{2}}\) 
 \\ & & & & \\
\(u=\mathcal{K}\) & \(1\) & \(0\) & \(k'_1\) & \(k'_2\) \\ & & & & \\
\(u=3/2\mathcal{K}\) & \((1+k'_1k'_2)^{-\frac{1}{2}}\) & \(-(k'_1k'_2)^{\frac{1}{2}}(1+k'_1k'_2)^{-\frac{1}{2}}\)& 
\({k'_1}^{\frac{1}{2}}(k'_1+k'_2)^{\frac{1}{2}}(1+k'_1k'_2)^{-\frac{1}{2}}\) & \({k'_2}^{\frac{1}{2}}(k'_1+k'_2)^{\frac{1}{2}}(1+k'_1k'_2)^{-\frac{1}{2}}\) 
\\ & & & & \\
\(u=i\mathcal{K}'/2\) & \(i(\kappa k_2'^2-k_2^2)^{-\frac{1}{2}}\) & \(\sqrt{\frac{\kappa k_2'^2-k_2^2+k_1^2}{\kappa k_2'^2-k_2^2}}\) &
\(\sqrt{1+\kappa}(1-\frac{k_2^2}{\kappa k_2'^2})^{-\frac{1}{2}}\) & \((1-\frac{k_2^2}{\kappa k_2'^2})^{-\frac{1}{2}}\) \\ & & & & \\
\(u=i\mathcal{K}'\) & \(k_2^{-1}\) & \(ik_2^{-1}k'_2\) & \(ik_2^{-1}(k_1^2-k_2^2)^{\frac{1}{2}}\) & \(0\) \\ & & & & \\
\(u=\mathcal{K}/2+i\mathcal{K}'/2\) & \(\sqrt{\frac{k_1^2+ik'_1k'_2\kappa}{k_1^2-k_2^2+k_1^2k_2^2}}\) &
\(\sqrt{\frac{-k_2^2{k'_1}^2-ik'_1k'_2\kappa}{k_1^2-k_2^2+k_1^2k_2^2}}\) & \(\sqrt{\frac{{k'_1}^2(k_1^2-k_2^2)-ik_1^2k'_1k'_2\kappa}{k_1^2-k_2^2+k_1^2k_2^2}}\) & 
\(\sqrt{\frac{k_1^2-k_2^2-ik_2^2k_1'k_2'\kappa}{k_1^2-k_2^2+k_1^2k_2^2}}\) \\ & & & & \\
\(u=\mathcal{K}/2+i\mathcal{K}'\) & \((1-k_1'k_2')^{-\frac{1}{2}}\) & \(-i(k_1'k_2')^{\frac{1}{2}}(1-k_1'k_2')^{-\frac{1}{2}}\) & 
\(-ik_1'^{\frac{1}{2}}(k_1'-k_2')^{\frac{1}{2}}(1-k_1'k_2')^{-\frac{1}{2}}\) & \(-ik_2'^{\frac{1}{2}}(k_1'-k_2')^{\frac{1}{2}}(1-k_1'k_2')^{-\frac{1}{2}}\) \\ & & & & \\
\(u=\mathcal{K}+i\mathcal{K}'\) & \(k_1^{-1}\) & \(ik_1^{-1}k_1'\) & \(0\) & \(k_1^{-1}(k_1^2-k_2^2)^{\frac{1}{2}}\) \\ 
\hline
\end{tabular}
\end{table}

The other values in Table II can be shown in similar ways using the addition theorems appropriately.
%\begin{figure}
% \includegraphics[scale=1]{HyperellK.ps}
% \caption{\(\mathcal{K}(k_1,k_2)\) as function of \(k_1\)}
%\end{figure} 

Together with the addition theorems one finds further
\begin{eqnarray}
 s(u+\mathcal{K})=\frac{c(u)}{\sqrt{d_1^2(u)-k_2^2{k'_1}^2s^2(u)}},\qquad c(u+\mathcal{K})=-\frac{k'_1k'_2s(u)}{\sqrt{d_1^2(u)-k_2^2{k'_1}^2s^2(u)}},
 \nonumber\\
 d_1(u+\mathcal{K})=\frac{k'_1d_2(u)}{\sqrt{d_1^2(u)-k_2^2{k'_1}^2s^2(u)}},\qquad d_2(u+\mathcal{K})=\frac{k'_2d_1(u)}{\sqrt{d_1^2(u)-k_2^2{k'_1}^2s^2(u)}}.
\end{eqnarray}

\subsection{The integrals of generalized Jacobi functions}

It is easy to see that the integral of \(d_2^2(u)\) is closely related to the incomplete elliptic integral of the third kind:
\begin{equation}
 \int\mathrm{d}ud_2^2(u)=\int\frac{\mathrm{d}u}{1+\frac{k_2^2}{k_2'^2}\mathrm{sn}^2(k_2'u,\kappa)}=\frac{1}{k_2'}\mathbf{\Pi}\left(k_2'u,
 -\frac{k_2^2}{k_2'^2},\kappa\right).
\end{equation}

Using (\ref{eq:Companion}) we get the corresponding integrals of \(s^2(u),c^2(u)\) and \(d_1^2(u)\), see Table III.

\begin{table}\label{tab:int}
\caption{A integral table of generalized Jacobi elliptic functions}
\begin{tabular}{c|c||c|c}
\hline
 f(u) & F(u) & f(u) & F(u)\\
 \hline\hline
 \(s(u)\) & \(-\frac{k'_2}{k_1}\mathrm{sn}^{-1}\left(\frac{k_1c(u)}{d_1(u)},\frac{\kappa}{k_2}\right)\) &
 \(s(u)d_2(u)\) & \(\frac{1}{k_1}\ln\left(d_1(u)-k_1c(u)\right)\)\\ & & & \\
 \(c(u)\) & \(\frac{1}{k_1}\mathrm{sn}^{-1}\left(k_1s(u),\frac{k_2}{k_1}\right)\) &
 \(c(u)d_1(u)\) & \(\frac{1}{k_2}\arctan\left(\frac{k_2s(u)}{d_2(u)}\right)\)\\  & & & \\
 \(d_1(u)\) & \(\mathrm{sn}^{-1}(s(u),k_2)\) &
 \(c(u)d_2(u)\) & \(\frac{1}{k_1}\arctan\left(\frac{k_1s(u)}{d_1(u)}\right)\)\\ & & & \\
 \(d_2(u)\) & \(\mathrm{sn}^{-1}(s(u),k_1)\) &
 \(d_1(u)d_2(u)\) & \(a(u)\) \\ & & & \\
 \(s(u)c(u)\) & \(\frac{1}{k_1k_2}\ln\left(k_2d_1(u)-k_1d_2(u)\right)\) &
 \(d_2^2(u)\) & \(\frac{1}{k_2'}\mathbf{\Pi}\left(k_2'u,-\frac{k_2^2}{k_2'^2},\kappa\right)\)\\ & & & \\
 \(s(u)d_1(u)\) & \(\frac{1}{k_2}\ln\left(d_2(u)-k_2c(u)\right)\) &
 \(d_1^2(u)\) & \(\left(1-\frac{k_1^2}{k_2^2}\right)u+
 \frac{1}{k_2'}\frac{k_1^2}{k_2^2}\mathbf{\Pi}\left(k_2'u,-\frac{k_2^2}{{k'_2}^2},\kappa\right)\)\\ & & & \\
 \(s^2(u)\) & \(\frac{1}{k_2^2}u-\frac{1}{k_2'k_2^2}\mathbf{\Pi}\left(k_2'u,-\frac{k_2^2}{{k'_2}^2},\kappa\right)\) &
 \(d_1^2(u)d_2^2(u)\) & \(\frac{k'_2}{2}\left[\mathbf{E}(k'_2u,\kappa)-
 \frac{k_1^2}{k_2^2}k'_2u+\left(\frac{1}{{k'_2}^2}+\frac{k_1^2}{k_2^2}\right)\mathbf{\Pi}\left(k'_2u,-\frac{k_2^2}{{k'_2}^2},\kappa\right)\right.+\)\\
 & & & \(+\left.\frac{k_2^2}{k'_2}\frac{s(u)c(u)d_1(u)}{d_2(u)}\right]\)\\ 
 \(c^2(u)\) & \(\frac{1}{k_2^2k'_2}\mathbf{\Pi}\left(k_2'u,-\frac{k_2^2}{k_2'^2},\kappa\right)-\frac{{k'_2}^2}{k_2^2}u\) & & \\
\hline 
\end{tabular}
\end{table}

\section{Generalized Jacobi functions as double sine-Gordon kinks}

We are now able to discuss the (quasi-) periodic kink solutions of the double sine-Gordon model (DSG) 
\begin{equation}
 \mathcal{L}=\frac{1}{2}\partial_{\mu}\phi\partial^{\mu}\phi-V(\phi)
\end{equation}
where the potential is given by 
\begin{equation}\label{eq:DSGPot}
 V(\phi)=\frac{\mu}{\beta^2}\cos(\beta\phi)-\frac{\lambda}{\beta^2}\cos\left(\frac{\beta}{2}\phi\right)+C.
\end{equation}
We will choose the constant \(C\) in order to set the minima of the potential to zero, which gives
\begin{eqnarray}
 C &=& \frac{\lambda-\mu}{\beta^2},\qquad (\mu,\lambda>0\ \text{ and } \frac{\lambda}{4\mu}>1) \text{ or } \mu<0, \lambda>0,\\
 C &=& \frac{1}{\beta^2}(\frac{\lambda^2}{8\mu}+\mu), \qquad \mu>0,\qquad \frac{|\lambda|}{4\mu}<1,\\
 C &=& -\frac{\lambda+\mu}{\beta^2},\qquad \mu,\lambda<0,
\end{eqnarray}
The signs of the different terms of (\ref{eq:DSGPot}) are chosen, so that for \(\mu\to 0\) and \(\lambda>0\) the potential reduces to
the sine-Gordon potential:
\begin{equation}
 V(\phi)\stackrel{\mu\to 0}{\longrightarrow}\frac{\lambda}{\beta^2}\left(1-\cos\left(\frac{\beta}{2}\phi\right)\right).
\end{equation}
The kinks are solutions of the first order equation of motion
\begin{equation}\label{eq:statickink}
 \frac{1}{2}\left(\frac{\mathrm{d}\phi}{\mathrm{d}x}\right)^2-V(\phi(x))=A,
\end{equation}
where \(A\) is some integration constant.
This model possesses a rich phase structure depending on the parameters \(\lambda\) and \(\mu\) \cite{Cond, Mus2}, e.g.
for \(\lambda/4\mu>1\) the only extrema of the potential are \(\phi=\frac{2\pi n}{\beta}\) with in particular \(\phi=0\) as minimum and
\(\phi=\frac{2\pi}{\beta}\) as maximum.

We will show in this section that the kink solutions and corresponding energy densities get a unique canonical expression in terms of 
generalized Jacobi functions.

By shifting \(\bar\phi(x)=\beta\phi(x)-2\pi\) the first order equation of motion for static kink configurations (\ref{eq:statickink}) 
can uniformly be brought to the form
\begin{equation}\label{eq:diffeq}
 \frac{\mathrm{d}\bar\phi}{\sqrt{(1-k_1^2\sin^2\frac{\bar\phi}{2})(1-k_2^2\sin^2\frac{\bar\phi}{2})}}=2\sqrt{\mu}\mathrm{d}x,
\end{equation} 
with solution 
\begin{equation}\label{eq:DSGkink}
 \phi(x)=\frac{2\pi}{\beta}+\frac{4}{\beta}a\left(\frac{\sqrt{\mu}}{k_1k_2}(x-x_0),k_1,k_2\right),
\end{equation}
which depends implicitly on the radius 
\begin{equation}\label{eq:parametricradius}
 R=\frac{2k_1k_2}{\sqrt{\mu}}\mathcal{K}(k_1,k_2).
\end{equation}
The corresponding energy density can be analytically expressed as
\begin{equation}\label{eq:DSGenden}
 \mathcal{E}(x,k_1,k_2,A)=\frac{16\mu}{\beta^2k_1^2k_2^2}d_1^2\left(\frac{\sqrt{\mu}}{k_1k_2}x,k_1,k_2\right)d_2^2
 \left(\frac{\sqrt{\mu}}{k_1k_2}x,k_1,k_2\right)-A,
\end{equation}
where \(d_1(x)\) and \(d_2(x)\) are the previous introduced generalized Jacobi functions.
(\ref{eq:DSGkink}) and (\ref{eq:DSGenden}) are the unique solution of the first order differential equation (\ref{eq:diffeq}).
The only thing one has to do, is to work out the explicit dependence of the moduli \(k_1,k_2\) on the parameters \(\mu,\lambda,\beta\) and the
integration constant \(A\) of the potential (\ref{eq:DSGPot}) in the different sectors. 
The solution has the following (quasi-)periodic properties, depending on the integration constant \(A\): 
\begin{eqnarray}
 \phi(x+R) &=& \phi(x)+\frac{4\pi}{\beta},\qquad A>0,\\
 \phi(x+2R) &=& \phi(x), \qquad A<0.
\end{eqnarray}
Depending on the physical situation these solutions can be used to describe kink chains on an infinite line or a kink solution 
on the compact circle with circumference \(R\).
Although (\ref{eq:DSGkink}) is 
in principle valid for all values of \(k_1,k_2\) we will give in addition for all cases an expression in text book functions where the
elliptic modulus \(\kappa\) lies in the fundamental interval between \(0\) and \(1\). This will establish the connection with previous obtained 
expressions for periodic solutions of the DSG model \cite{Iwab, Huda, Wang}.

\subsection{Case: \(\lambda,\mu >0\) and \(\lambda>4\mu\)}

In this region of the parameter space the potential (\ref{eq:DSGPot}) has only one type of minima. 
The moduli are given by
\begin{equation}\label{eq:kdependent}
 k_{1,2}^2=\frac{1}{\beta^2A+2\lambda}\left[4\mu+\lambda\pm\sqrt{(\lambda-4\mu)^2-8\mu\beta^2A}\right],
\end{equation}
with following properties
\begin{equation}
 k_1^2k_2^2=\frac{8\mu}{\beta^2A +2\lambda},\qquad k_1^2+k_2^2=\frac{8\mu+2\lambda}{\beta^2A+2\lambda},\qquad 
 k_1^2+k_2^2-k_1^2k_2^2=\frac{2\lambda}{\beta^2A+2\lambda},\qquad
 {k'_1}^2{k'_2}^2=\frac{\beta^2A}{\beta^2A+2\lambda}.
\end{equation}

\begin{figure}
 \includegraphics[scale=0.5]{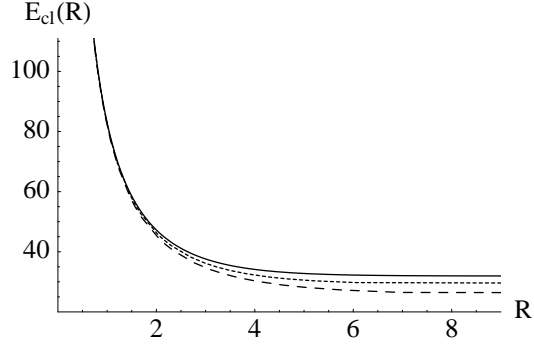}
 \caption{Classical energy for \(\lambda=4\) and \(\mu=0.01\) (solid), \(\mu=0.5\) (dotted) and \(\mu=0.99\) (dashed)}
\end{figure}

\subsubsection{\(0<A<(\lambda-4\mu)^2/(8\mu)\)}\label{ch:subcaseA}
 In this case is \(0<k_2^2<k_1^2<1\) and the solution can be written in terms of elementary functions as
\begin{equation}
 \phi(x)=\frac{2\pi}{\beta}+\frac{4}{\beta}\arctan\left[{k'_2}^{-1}\mathrm{sc}\left(\frac{k'_2\sqrt{\mu}}{k_1k_2}(x-x_0),\kappa\right)\right],
\end{equation}
depending on the radius 
\begin{equation}
 R=\frac{2k_1k_2}{\sqrt{\mu}}\frac{1}{k'_2}\mathbf{K}(\kappa).
\end{equation}

This solution can be interpreted as an infinite kink chain on the line with distance \(R\).
The energy of this field configuration on \(S^1\) is
\begin{equation}\label{eq:parametricenergy}
 E(k_1,k_2)=\frac{16\sqrt{\mu}k'_2}{\beta^2k_1k_2}\left[\mathbf{E}(\kappa)-\left(\frac{k_1^2}{k_2^2}+1-k_1^2\right)\mathbf{K}(\kappa)+
 \left(\frac{1}{{k'_2}^2}+\frac{k_1^2}{k_2^2}\right)\mathbf{\Pi}\left(-\frac{k_2^2}{{k'_2}^2},\kappa\right)\right].
\end{equation}
With (\ref{eq:kdependent}) the radius (\ref{eq:parametricradius}) and energy (\ref{eq:parametricenergy}) become functions of \(A,\beta,\mu\) and \(\lambda\) 
\begin{equation}
 R=R(A;\beta,\lambda,\mu),\qquad E=E(A;\beta,\lambda,\mu),
\end{equation}
which can for given \(\beta,\lambda,\mu\) be plotted with parameter \(A\) (see Figure 5).

\begin{figure}
 \includegraphics[scale=0.7]{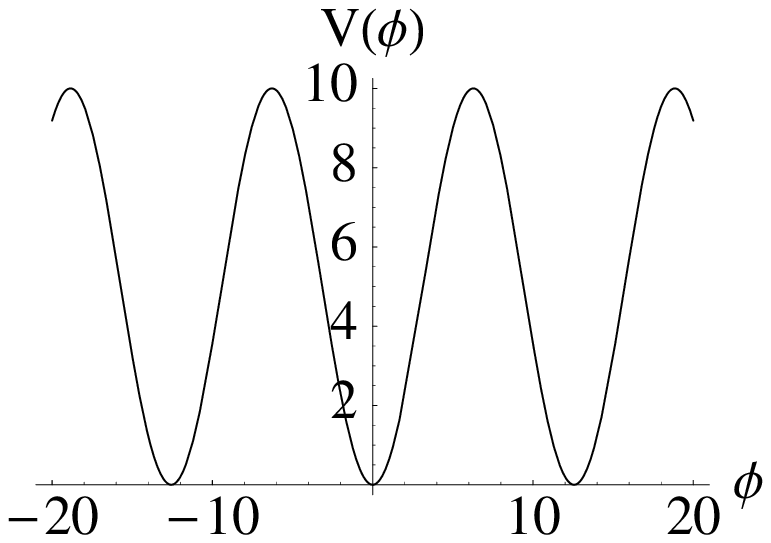}
 \includegraphics[scale=0.7]{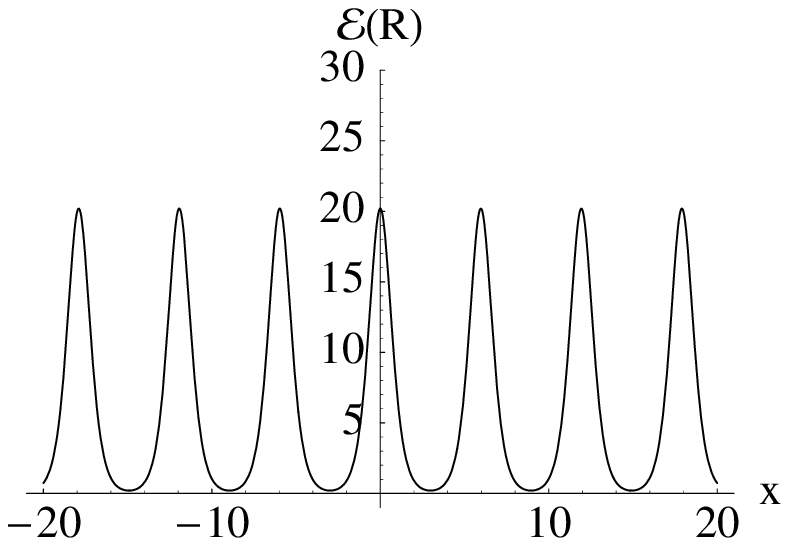}
 \caption{The DSG potential \(V(\phi)\) and the kink-chain energy density \(\mathcal{E}(x)\) for \(\lambda>4\mu\)}
\end{figure}

\subsubsection{\(A=0\)} 

This is the decompactification limit since 
from \(k_1^2=1\) and \(k_2^2=4\mu/\lambda\)  and (\ref{eq:parametricradius}) follows \(R\to\infty\).
Then the kink solution reduces to a single DSG kink on the infinite line
\begin{equation}
 \phi(x)\to \frac{2\pi}{\beta}+\frac{4}{\beta}\arctan\left[\sqrt{\frac{\lambda}{\lambda-4\mu}}\sinh
 \left(\sqrt{\frac{\lambda}{4}-\mu}(x-x_0)\right)\right],
\end{equation}
which is the solution found in \cite{Mus2}. The corresponding topological charge \(Q=\phi(+\infty)-\phi(-\infty)\) is
\begin{equation}
 Q=\frac{4\pi}{\beta}
\end{equation}

\subsubsection{\(A=(\lambda-4\mu)^2/(8\mu)\)}

This is the trigonometric point, since the moduli are given by
\begin{equation}
 k_1^2=k_1^2=k^2=\frac{8\mu}{\lambda+4\mu}.
\end{equation}
and in the kink solution all elliptic functions degenerate to trigonometric functions:
\begin{equation}
 \phi(x)\to\frac{2\pi}{\beta}+\frac{4}{\beta}\arctan\left[\frac{1}{k'}\tan\left(\frac{k'}{k^2}\sqrt{\mu}(x-x_0)\right)\right],\qquad
 R\to \frac{\pi}{\sqrt{\mu}}\frac{k^2}{k'}.
\end{equation}
The energy is 
\begin{equation}
 E\to \frac{4\pi\sqrt{\mu}}{\beta^2}\left[\frac{2-k^2}{k^2}+2\frac{2-k^2}{k'+1}-k^2\right].
\end{equation}

\begin{itemize}
\item \(\mu=0\)

This is the sine-Gordon limit

From (\ref{eq:kdependent}) one can see \(k_2\to 0\) and the quasi-periodic sine-Gordon soliton is obtain with
\begin{equation}
 \phi(x)\to\frac{2\pi}{\beta}+\frac{4}{\beta}\mathrm{am}\left(\frac{\sqrt{\lambda/4}}{k_1}(x-x_0),k_1\right)
\end{equation}
and
\begin{equation}
 R\to\frac{2k_1}{\sqrt{\lambda/4}}\mathbf{K}(k_1)
\end{equation}
with mass parameter \(m=\sqrt{\lambda/4}\).
By using the limit
\begin{equation}
 \lim_{k_2\to 0}\frac{k_1^2}{k_2^2}\left[\mathbf{\Pi}\left(-\frac{k_2^2}{{k'_2}^2},\kappa\right)-
 \mathbf{K}(\kappa)\right]=\mathbf{E}(k_1)-\mathbf{K}(k_1),
\end{equation}
the energy becomes
\begin{equation}
 E(k_1,k_2)\to\frac{16\sqrt{\lambda/4}}{\beta^2k_1}\left[(k_1^2-1)\mathbf{K}(k_1)+2\mathbf{E}(k_1)\right]
\end{equation}
\end{itemize}

\subsubsection{\(A>(\lambda-4\mu)^2/(8\mu)\)}

Now the moduli \(k_1^2\) and \(k_2^2\) are complex conjugated  with
\begin{equation}
 |k_1^2|^2=|k_2^2|^2=\frac{8\mu}{A+2\mu}
\end{equation}
The explicit kink solution can be written as
\begin{equation}
 \phi(x)=\frac{2\pi}{\beta}+\frac{4}{\beta}\arctan\left[(k'_1k'_2)^{-1/2}\mathrm{sc}\left(
 \frac{\sqrt{k'_1k'_2}\sqrt{\mu}}{k_1k_2}x,i\frac{(k'_1-k'_2)}{2\sqrt{k'_1k'_2}}\right)\mathrm{dn}\left(
 \frac{\sqrt{k'_1k'_2}\sqrt{\mu}}{k_1k_2}x,i\frac{(k'_1-k'_2)}{2\sqrt{k'_1k'_2}}\right)\right],
\end{equation}
where the radius is given by
\begin{equation}
 R=\frac{2k_1k_2}{\sqrt{\mu}}\frac{1}{\sqrt{k'_1k'_2}}\mathbf{K}\left(i\frac{(k'_1-k'_2)}{2\sqrt{k'_1k'_2}}\right)
\end{equation} 
This is again a kink chain as in Case A.1, only the mathematical representation has changed. 

\subsubsection{\(-2\lambda<A<0\)}

In this case \(0<k_2^2<1<k_1^2<\infty\) and the solution can be written as
\begin{equation}
 \phi(x)=\frac{2\pi}{\beta}+\frac{4}{\beta}\arctan\left[(k_1^2-k_2^2)^{-1/2}\mathrm{sd}\left(\frac{\sqrt{k_1^2-k_2^2}\sqrt{\mu}}{k_1k_2}x,
 \kappa^{-1}\right)\right],
\end{equation}
where the radius is given by
\begin{equation}
 R=\frac{2k_1k_2}{\sqrt{\mu}}\frac{1}{\sqrt{k_1^2-k_2^2}}\mathbf{K}(\kappa^{-1})
\end{equation}
This solution can be interpreted as an infinite chain of kinks and anti-kinks on the line.
 
\subsubsection{\(A=-2\lambda\)}

This is the endpoint for real valued solutions in the DSG model, where the moduli become
\begin{equation}
 k_1^2\to\infty,\qquad k_2^2=(1-\frac{\lambda}{4\mu})^{-1},
\end{equation}
and the kink solution reduces the constant field configuration 
\begin{equation}
 \phi(x)=\frac{2\pi}{\beta},
\end{equation}
with constant energy density 
\begin{equation}
 \mathcal{E}(x)=\frac{2\lambda}{\beta^2}.
\end{equation}
This happens at the critical value
\begin{equation}\label{eq:critrad}
 R_0=\frac{2\pi}{\sqrt{4\mu+\lambda}}.
\end{equation}
Thus for \(R<R_0\) no non-trivial real valued periodic static field configuration exist in the DSG model.

\subsection{Case: \(|\lambda|<4\mu\) and \(\mu>0\) }

The potential (\ref{eq:DSGPot}) has now two different maxima and additional minima.

The kink solution is again (\ref{eq:DSGkink}) with the moduli given by
\begin{equation} 
 k_{1,2}^2=\frac{1}{\beta^2A+\frac{1}{8\mu}(\lambda+4\mu)^2}\left[4\mu+\lambda\pm\sqrt{-8\mu\beta^2A}\right].
\end{equation}

Since the DSG potential (\ref{eq:DSGPot}) has the symmetry
\begin{equation}
 V(\phi,\lambda)=V\left(\phi+\frac{2\pi}{\beta},-\lambda\right)
\end{equation}
the second solutions are given by
\begin{equation}
 \phi_{II}(x,\lambda,A)=\phi_{I}(x,-\lambda,A)-\frac{2\pi}{\beta}.
\end{equation}

\subsubsection{\(A>0\)}

The moduli are complex conjugated with
\begin{equation}
 |k_1^2|^2=|k_2^2|^2=\frac{8\mu}{\beta^2A+\frac{1}{8\mu}(\lambda+4\mu)^2},
\end{equation}
 
and the kink solution is 
 \begin{equation}\label{eq:DSGkink3}
 \phi_I(x)=\frac{2\pi}{\beta}+\frac{4}{\beta}\arctan\left[(k'_1k'_2)^{-1/2}\mathrm{sc}\left(
 \frac{\sqrt{k'_1k'_2}\sqrt{\mu}}{k_1k_2}x,i\frac{(k'_1-k'_2)}{2\sqrt{k'_1k'_2}}\right)\mathrm{dn}\left(
 \frac{\sqrt{k'_1k'_2}\sqrt{\mu}}{k_1k_2}x,i\frac{(k'_1-k'_2)}{2\sqrt{k'_1k'_2}}\right)\right],
\end{equation}
where the radius is given by
\begin{equation}
 R_I=\frac{2k_1k_2}{\sqrt{\mu}}\frac{1}{\sqrt{k'_1k'_2}}\mathbf{K}\left(i\frac{(k'_1-k'_2)}{2\sqrt{k'_1k'_2}}\right).
\end{equation} 
On \(S^1\) this solution represents a quasi-periodic kink. On the infinite line this solution represents a chain composed 
of two different types of kinks, a large and a small one, where the large kink lies around \(x=0\). This can be seen on the energy density 
chart.
The second solution \(\phi_{II}(x)\) is equivalent to \(\phi_I(x)\), but now the small kink lies around \(x=0\).

\subsubsection{\(A=0\)}
 
 The moduli are \(k_{1,2}^2=\frac{8\mu}{4\mu+\lambda}\) and \(R\to\infty\).
 The solution \(I\) reduces for \(\lambda>0\) to the single large kink:
 \begin{equation}
  \phi_I(x)=\frac{2\pi}{\beta}+\frac{4}{\beta}\arctan\left[\sqrt{\frac{4\mu+\lambda}{4\mu-\lambda}}\tanh\left(\sqrt{
  1-\left(\frac{\lambda}{4\mu}\right)^2}\frac{\sqrt{\mu}}{2}x\right)\right],
 \end{equation}
and for \(\lambda<0\) to the small kink.
Solution \(II\) gives for \(\lambda>0\) the single small kink and for \(\lambda<0\) the single large kink:
\begin{equation}
 \phi_{II}(x)=\frac{4}{\beta}\arctan\left[\sqrt{\frac{4\mu-\lambda}{4\mu+\lambda}}\tanh\left(\sqrt{
  1-\left(\frac{\lambda}{4\mu}\right)^2}\frac{\sqrt{\mu}}{2}x\right)\right]
\end{equation}
The corresponding topological charge is given by
\begin{equation}
 Q_{I,II}=\frac{8}{\beta}\arctan\left[\sqrt{\frac{4\mu\pm\lambda}{4\mu\mp\lambda}}\right]
\end{equation}
The obvious relation \(Q_I>Q_{II}\) for \(\lambda>0\) justifies the nomenclature large/small kink. 
\begin{figure}
 \includegraphics[scale=0.7]{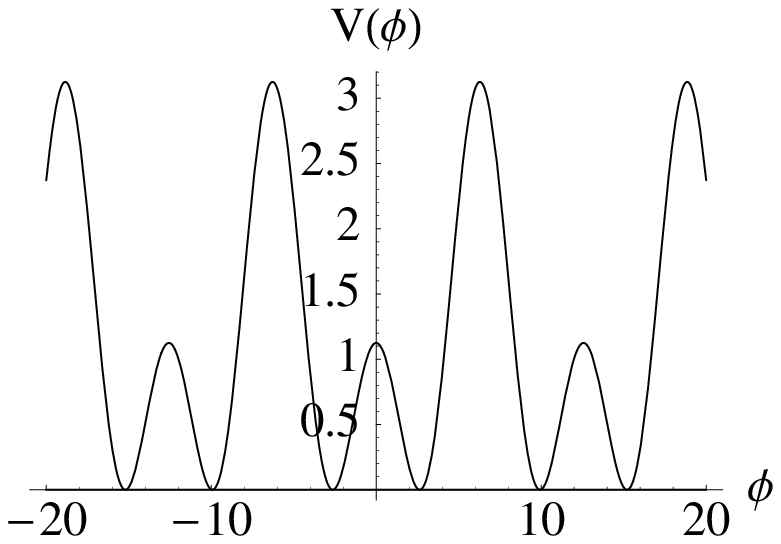}
 \includegraphics[scale=0.7]{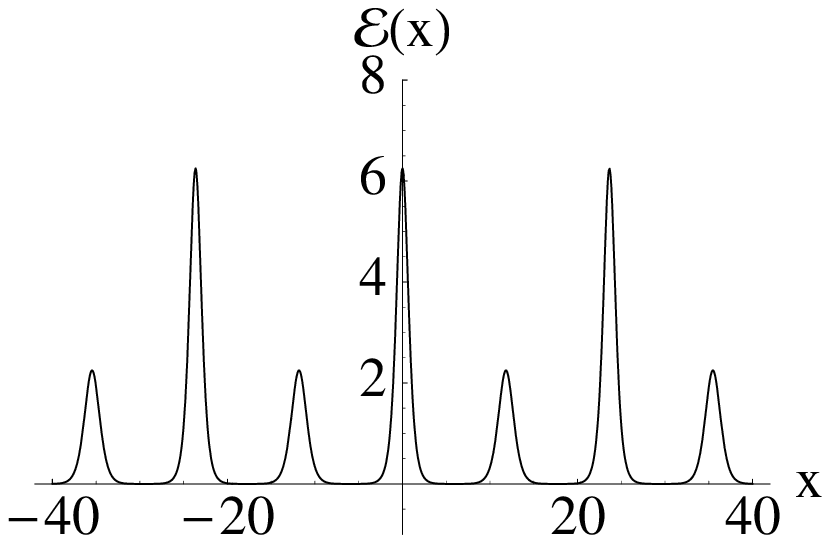}
 \caption{The DSG potential \(V(\phi)\) and energy density \(\mathcal{E}(x)\) of a chain of large-kinks/small-kinks for \(\lambda<4\mu\)}
\end{figure}

\subsubsection{\(-(\lambda-4\mu)^2/(8\mu)<A<0\)}

The moduli are real with \(1<k_2^2<k_1^2<\infty\). Now there are two inequivalent solutions. The first one can now be written as
\begin{equation}
 \phi_I(x)=\frac{2\pi}{\beta}+\frac{4}{\beta}\arctan\left[(k_1^2-1)^{-1/2}\mathrm{sn}\left(\frac{\sqrt{k_1^2-1}\sqrt{\mu}}{k_1k_2}x,\
 \kappa'^{-1}\right)\right],\qquad R=\frac{2k_1k_2}{\sqrt{\mu}\sqrt{k_1^2-1}}\mathbf{K}\left(\kappa'^{-1}\right).
\end{equation}
The second one is
\begin{equation}
 \phi_{II}(x,\lambda,A)=\phi_{I}(x,-\lambda,A)-\frac{2\pi}{\beta}.
\end{equation}
Solution \(\phi_I\) represents for \(\lambda>0\) a chain of kinks and anti-kinks of the 
large type with distance \(R\) and for \(\lambda<0\)  a chain of kinks and anti-kinks of 
the small type with distance \(R\).

\subsubsection{\(\lambda<0\) and \(A=-(\lambda-4\mu)^2/(8\mu)\)}

This is the endpoint of the kink/anti-kink chain of the small type. For the moduli we have
\begin{equation}
 k_1^2\to\infty,\qquad k_2^2=\frac{4\mu}{\lambda+4\mu},
\end{equation}
and the kink reduces to the constant field configuration
\begin{equation}
 \phi(x)=\frac{2\pi}{\beta},
\end{equation}
with energy density 
\begin{equation} 
 \mathcal{E}(x)=\frac{2\mu}{\beta^2}\left(1+\frac{\lambda}{4\mu}\right)^2.
\end{equation}
This happens at the critical value 
\begin{equation}
 R_0=\frac{2\pi}{\sqrt{-|\lambda|+4\mu}}
\end{equation} 

\subsubsection{\(\lambda>0\) and \(-2\mu(1+\lambda/(4\mu))^2< A<-(\lambda-4\mu)^2/(8\mu)<0\)}

In this case \(0<k_2^2<1<k_1^2<\infty\) and the solution can be written as
\begin{equation}
 \phi(x)=\frac{2\pi}{\beta}+\frac{4}{\beta}\arctan\left[(k_1^2-k_2^2)^{-1/2}\mathrm{sd}\left(\frac{\sqrt{k_1^2-k_2^2}\sqrt{\mu}}{k_1k_2}x,
 \kappa^{-1}\right)\right],\qquad
 R=\frac{2k_1k_2}{\sqrt{\mu}}\frac{1}{\sqrt{k_1^2-k_2^2}}\mathbf{K}(\kappa^{-1}).
\end{equation}
This is the kink/anti-kink chain of the large type.
\subsubsection{\(\lambda>0\) and \(A=-2\mu(1+\lambda/(4\mu))^2\)}

This is the endpoint of the kink/anti-kink chain of the large type. For the moduli we have
\begin{equation}
 k_1^2\to\infty,\qquad k_2^2=\frac{4\mu}{\lambda+4\mu},
\end{equation}
and the kink reduces to the constant field configuration
\begin{equation}
 \phi(x)=\frac{2\pi}{\beta}
\end{equation}
at the critical radius \(R_0\) given by (\ref{eq:critrad}). 

\subsection{Case: \(\lambda,\mu <0\) and \(A>0\)}

The potential (\ref{eq:DSGPot}) has now two different minima.
The kink is again (\ref{eq:DSGkink}) with the following moduli:
\begin{equation}
 k_{1,2}^2=\frac{1}{\beta^2A}\left[4\mu+\lambda\pm\sqrt{(\lambda-4\mu)^2+8\mu(2\lambda-\beta^2A)}\right],
\end{equation}
with
\begin{equation}
 k_2^2<-1<0<k_1^2<1.
\end{equation}
Therefore an explicit representation of the kink in terms of text book functions is 
\begin{equation}
 \phi(x)=\frac{2\pi}{\beta}+\frac{4}{\beta}\arctan\left[{k'_2}^{-1}\mathrm{sc}\left(\frac{k'_2\sqrt{-\mu}}{k_1\sqrt{-k_2^2}}(x-x_0),\kappa\right)\right],
 \qquad  R=\frac{2k_1\sqrt{-k_2^2}}{k'_2\sqrt{-\mu}}\mathbf{K}(\kappa).
\end{equation}
On the infinite line one can interpret this as a chain of two small kinks bounded in a kind of molecule, see Figure 9.

\begin{figure}
 \includegraphics[scale=0.7]{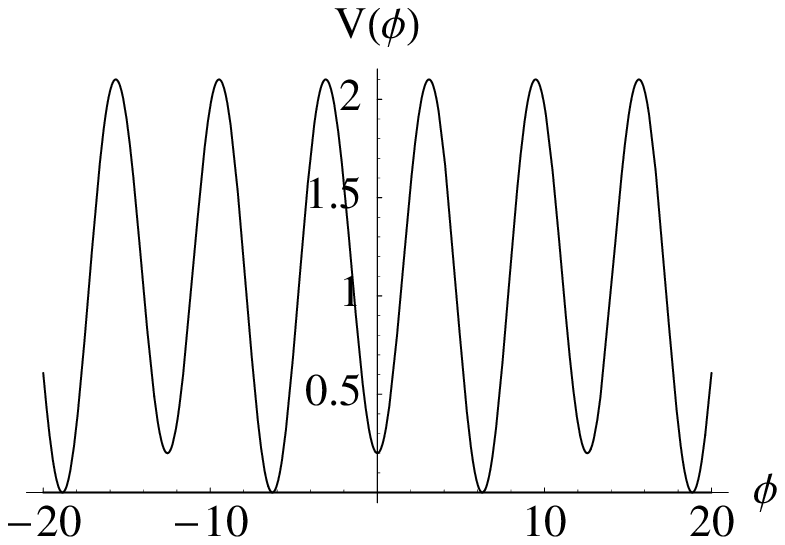}
 \includegraphics[scale=0.7]{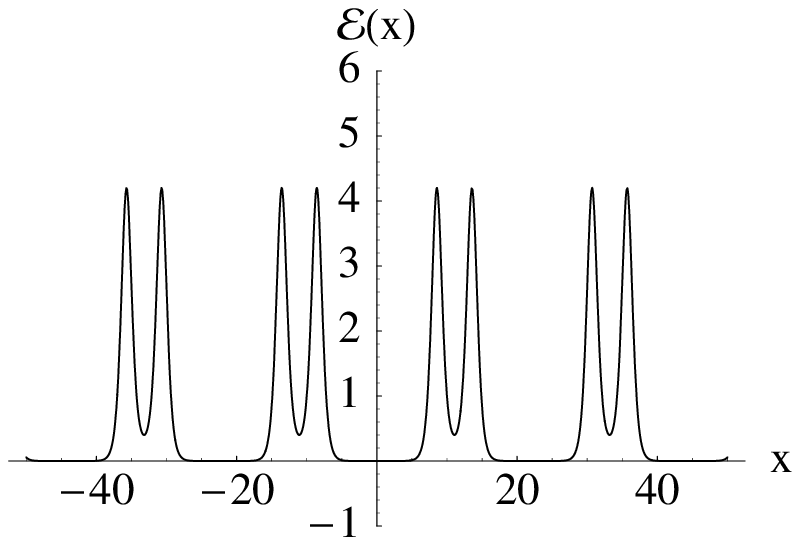}
 \caption{The DSG potential \(V(\phi)\) and energy density \(\mathcal{E}(x)\) of a chain of kink-kink molecules for \(\lambda,\mu<0\)}
\end{figure}

\subsection{Case: \(\mu<0\) and \(\lambda>0\)}

The moduli are the same as for case A: 
\begin{equation}
 k_{1,2}^2=\frac{1}{\beta^2A+2\lambda}\left[4\mu+\lambda\pm\sqrt{(\lambda-4\mu)^2-8\mu\beta^2A}\right],
\end{equation}

\subsubsection{\(A>0\)}
The moduli are \(k_2^2<0<k_1^2<1\). The kink solution is
\begin{equation}
 \phi(x)=\frac{2\pi}{\beta}+\frac{4}{\beta}\arctan\left[{k'_2}^{-1}\mathrm{sc}\left(\frac{k'_2\sqrt{-\mu}}{k_1\sqrt{-k_2^2}}(x-x_0),\kappa
 \right)\right], \qquad R=\frac{2k_1\sqrt{-k_2^2}}{\sqrt{-\mu}}\frac{1}{k'_2}\mathbf{K}(\kappa).
\end{equation}
 
\subsubsection{\(\lambda>4|\mu|\) and \(-2\lambda<A<0\)}
The moduli are \(k_2^2<0<1<k_1^2\). The kink solution is
\begin{equation}
 \phi(x)=\frac{2\pi}{\beta}+\frac{4}{\beta}\arctan\left[(k_1^2-k_2^2)^{-1}\mathrm{sd}\left(\frac{\sqrt{k_1^2-k_2^2}}{k_1\sqrt{-k_2^2}}
 \sqrt{-\mu}(x-x_0),\kappa^{-1}\right)\right],\qquad R=\frac{2k_1\sqrt{-k_2^2}}{\sqrt{-\mu}}\frac{1}{\sqrt{k_1^2-k_2^2}}\mathbf{K}(\kappa^{-1}).
\end{equation}
This is a periodic bounce solution.

\section{Conclusion}
We introduced a generalization of Jacobi elliptic functions defined by the inversion of certain hyperelliptic integrals which
are reducible to elliptic integrals. 

As an example for their effectiveness in physics we
have chosen the double sine-Gordon model. Its (quasi-)periodic kink solution and corresponding energy densities can be described uniformly 
by a single generalized Jacobi function. The qualitative characteristics of the kink chains depend only on the moduli parameter \(k_1\) and 
\(k_2\). Several solutions of the DSG model obtained in the past \cite{Iwab, Huda, Wang} are just special cases of a unique generalized Jacobi
function. We observed also a critial value \(R_0\) for kink/anti-kink chains, where for \(R<R_0\) no non-trivial static solution exists.

\end{document}